\documentclass{eptcs}
\usepackage{import}
\usepackage{booktabs}
\usepackage{ragged2e}
\usepackage{setspace}
\usepackage{lipsum} % to generate text
\usepackage{titlesec}
\usepackage{fancyhdr}
\usepackage{lastpage}

\usepackage{amsmath}
\usepackage{amssymb}
\usepackage{pifont}
\usepackage{textcomp}
\usepackage{xfrac}
\usepackage{setspace}
\usepackage{verbatim}
\usepackage{hyperref}

\newcommand{\cmark}{\ding{51}}%
\newcommand{\xmark}{\ding{55}}%

\usepackage{amsmath}
\usepackage{graphicx}
\usepackage{multirow}

\providecommand{\urlalt}[2]{\href{#1}{#2}}
\providecommand{\doi}[1]{doi:\urlalt{http://dx.doi.org/#1}{#1}}

\begin{document}

%\lefttitle{Maxime Mahout}

%\jnlPage{1}{14}
%\jnlDoiYr{2026}
%\doival{10.1017/xxxxx}

\title{A ProbLog program to infer individual genotypes from familial phenotypes in autosomal, X-linked, and Y-linked Mendelian disorders}

\newcommand{\titlerunning}[1]{\def\titlerunning{#1}}
\newcommand{\authorrunning}[1]{\def\authorrunning{#1}}

\titlerunning{A ProbLog program for Mendelian inheritance analyses}
\authorrunning{Maxime Mahout}

%\begin{authgrp}
\author{{Maxime} {Mahout}}

%\end{authgrp}

%\history{\sub{xx xx xxxx;} \rev{xx xx xxxx;} \acc{xx xx xxxx}}

\maketitle

\begin{abstract}
The automated reconstruction of patient family history is a common challenge in genetic counseling for disease prevention. Such a family history is usually determined for a particular subset of diseases that are Mendelian, \textit{i.e.} monogenic, and classified into three categories depending on the chromosome the gene is located: autosomal, X-linked or Y-linked. Mendel’s inheritance laws allow for simple probabilistic modeling of the genetic transmission of monogenic disorders. Genetic counsellors use knowledge about the patient’s family history and Mendelian laws for assessing risks of transmitting or inheriting congenital conditions. We present \textit{mendelprob.pl}, a probabilistic logic programming algorithm in ProbLog for deriving probabilities of inheritance of genotypes and phenotypes for genes with two alleles through multiple generations. In particular, the user can input genotypes and phenotypes for a patient and its family, and automatically determine the most probable genetic family history. We illustrate the ProbLog model on practical examples of patient pedigrees from the literature and from a genetic counseling handbook. We show that our method correctly infers probability of individual genotypes from knowledge about familial genotypes, yielding the same results as tool \textit{pedprobr}. However, unlike \textit{pedprobr}, our approach can exploit knowledge about familial phenotypes. It can also directly distinguish between autosomal, X-linked, and Y-linked disorders, using its intuitive logical modelling. We provide our ProbLog tool for free and open-source on GitHub, making it easily available for genetic counsellors. We conclude on the importance of providing explainable formal methods for a task that clinicians might want to perform using proprietary software.
\end{abstract}

%\begin{keywords}
\vspace{5pt}

Correspondence: {\href{email:maxime.mahout@inria.fr}{maxime.mahout@inria.fr}. OrcId: \href{https://orcid.org/0000-0002-2699-5582}{0000-0002-2699-5582}}. 

Affiliation: {{Inria Saclay}, {EPI Lifeware}, {{91120}, {Palaiseau}, {France}}}.

Keywords: Probabilistic logic programming, Classical genetics, Logical inference.
%\end{keywords}

\section*{Introduction}

In 1865, Gregor Mendel first described on species of peas the notion of \textit{dominant} and \textit{recessive} characteristics, characteristics that might be either predominantly conserved, or predominantly lost in a hybrid \cite{mendel_versuche_1865}. With his work, Mendel described what would later become known as alleles, genotypes and phenotypes. A genotype, the set of all alleles -- versions of genes of an individual -- is responsible for the phenotype -- the set of all observable characteristics of the individual. The notions of genotype and phenotype are meant to be observed together for two or more individuals of the same species \cite{orgogozo_differential_2015}. Mendel's legacy is still carried today through the concept of Mendelian diseases \cite{botstein_discovering_2003}. Modern genetics distinguish monogenic conditions, which obey Mendel's laws, and for which the responsible allele is confined to a single gene locus, from polygenic conditions, in which a condition can be caused by the association of multiple alleles at different loci. For polygenic conditions, the rise of genome-wide association studies has permitted significant advances in the medical treatment and diagnosis of illnesses \cite{visscher_five_2012}. Nevertheless, genome-wide association studies can sometimes overcomplicate analysis, and with their ease of availability in the next-generation sequencing era, not enough emphasis has been put on monogenic or quasi-monogenic diseases \cite{antonarakis_mendelian_2006, tam_benefits_2019}. %\vspace{25pt} %\pagebreak[4]  

Genetic counseling for disease prevention typically involves screening patients' family history for monogenic diseases \cite{gordon_future_2018, bennett_practical_2011}. Mendelian disorders are reported in online databases such as OMIM (Online Mendelian Inheritance in Man) and OrphaNet \cite{hamosh_online_2005, pavan_clinical_2017}. They are classified into five disease types, indicating which chromosome the responsible gene is located on, between autosomes, X and Y, and if the mutant allele is {dominant} or {recessive}: \textit{autosomal recessive}, \textit{autosomal dominant}, \textit{X-linked recessive}, \textit{X-linked dominant}, and \textit{Y-linked}. Genetic family history is often looked at using proprietary software such as MeTree \cite{ginsburg_family_2019}. These softwares are typically closed-source and do not involve logical reasoning; they need to be provided with prior knowledge of the patient's condition. There is therefore an interest in providing in such programs a way to directly infer genotypes and phenotypes, or to infer whether the phenotypes of the family indicate an X-linked, Y-linked or autosomal disorder.

In this study we present \textit{mendelprob.pl}, a probabilistic program in ProbLog that is able to simulate Mendelian inheritance for all five described allele transmission types. We show that we can logically derive conclusions on the genotypes and phenotypes of a pedigree for any potentially unknown monogenic disease. We illustrate that our tool can help derive genetic family history for probands, \textit{i.e.} people who seek genetic counseling. As well, we show that the tool's predictive performance for analyzing Mendelian inheritance of human disorders is on par with the pedigree genotyping method \textit{pedprobr} \cite{vigeland_pedigree_2021}. Our ProbLog tool's major advantage over \textit{pedprobr} is that -- thanks to probabilistic logic -- it does not require to have prior knowledge about the expected Mendelian disorder. Particularly, it can infer a patient's genotype from its phenotype, and it can determine what type of disorder is most likely associated to the genetic family history between X-linked, Y-linked and autosomal.

\section*{Methods}

Probabilistic programming is a well-established field of artificial intelligence, with many recent developments. Probabilistic logic programming is a subset of this field that proposes to use automated logical reasoning to derive probabilities for modeling uncertain events. The principle is as follows: probabilistic events are described with logic rules weighted by probabilities, and the solver's role is to find an assignment respecting truth values and the events' probability distribution. A probabilistic logic program is such a set of probabilistic logic rules and facts, which can be used to derive the probability of natural language queries. Probabilistic logic programming problems can be solved using solvers such as ProbLog \cite{de_raedt_problog_2007}. The tool relies on knowledge compilation techniques such as Binary Decision Diagrams \cite{bryant_graph-based_1986}: with the diagram's edges being weighted by probabilities, solutions are retrieved as tree traversals \cite{de_raedt_problog_2007}. For our application study, we are using the second version of ProbLog, which allows for setting evidence on atoms and multiple querying \cite{dries_problog2_2015, fierens_inference_2015}. Our tool, which we called \textit{mendelprob.pl}, is compared to the R tools \textit{pedsuite} by Vigeland \cite{vigeland_pedigree_2021}. All analyses and code are available at \href{https://github.com/maxm4/mendelprob.pl}{https://github.com/maxm4/mendelprob.pl}. % ProbLog is ran with the options \texttt{--knowledge-compilation=ddnnf} \texttt{--dont-propagate-evidence} for optimal running times.

\subsection*{Phenotype and genotype inference}

A key component of our \textit{mendelprob.pl} tool is its ability to derive probable genotypes from phenotypes and vice versa. Precisely, the tool applies to monogenic disorders that can be represented as bi-allelic, that is, disorders that can be modeled as a single gene with two states: such as healthy/wild type and ill/mutated. Let us call the wild type and mutant alleles respectively \textit{wt} and \textit{m}. From there, in classical genetics, the notions of \textit{dominance} and \textit{recessiveness} come into play. As traditionally, we denote the recessive allele $a$ and the dominant allele $A$. These notions explain whether the phenotype is only visible in homozygotes, for \textit{recessive} alleles (\textit{aa}) or if it is also visible in heterozygotes, for \textit{dominant} alleles (\textit{Aa}, \textit{AA}). Prevalence is the occurrence of an allele, usually a disorder-causing mutation, in the general population. We provide a way to modulate the prevalence $p$ of $a$, and prevalence of $A$ is given by $1-p$. %\pagebreak[4]
%As is standard, we may denote two possible zygote alleles \textit{a} for a recessive one and \textit{A} for a dominant one.

A disorder caused by a mutation $m$ can generally be either \textit{dominant} or \textit{recessive}. If it is recessive, then $m = a$ and $wt = A$, and the prevalence $p$ is the prevalence of $m$. If it is dominant, then $m = A$ and $wt = a$, and $1 - p$ is the prevalence of $m$. Therefore we can model genotypes and phenotypes of individuals for dominant and recessive monogenic disorders using only alleles \textit{a} and \textit{A}. In particular, for an autosomal disorder, the phenotypes are noted as $a$ if the person's genotype is $aa$, and $A$ if person's genotype is $Aa$ or $AA$. For sex-linked disorders, \textit{i.e.} X-linked and Y-linked, five additional genotypes are possible: \textit{-}, \textit{a} or \textit{A} for Y-linked, and \textit{a-} and \textit{A-} for males with a X-linked disorder.

In the logic program code, genotypes are represented by \texttt{carry} predicates, and phenotypes by \texttt{show} predicates. Since genotype inheritance probabilities are different between males and females in sex-linked disorders, the logic atoms are specified into \texttt{f\_carry}, \texttt{m\_carry}, \texttt{f\_show}, \texttt{m\_show}. The \texttt{show} atoms can receive $a$ the recessive phenotype or $A$ the dominant phenotype. The \texttt{carry} atoms can receive the eight aforementioned genotypes. As a feature of ProbLog, probabilities of genotypes and phenotypes are queried using \texttt{query} commands. Prior knowledge can also be integrated to specify these atoms using \texttt{evidence} commands.

\begin{figure*}[!t]%
\centering
\includegraphics[width=0.9\textwidth]{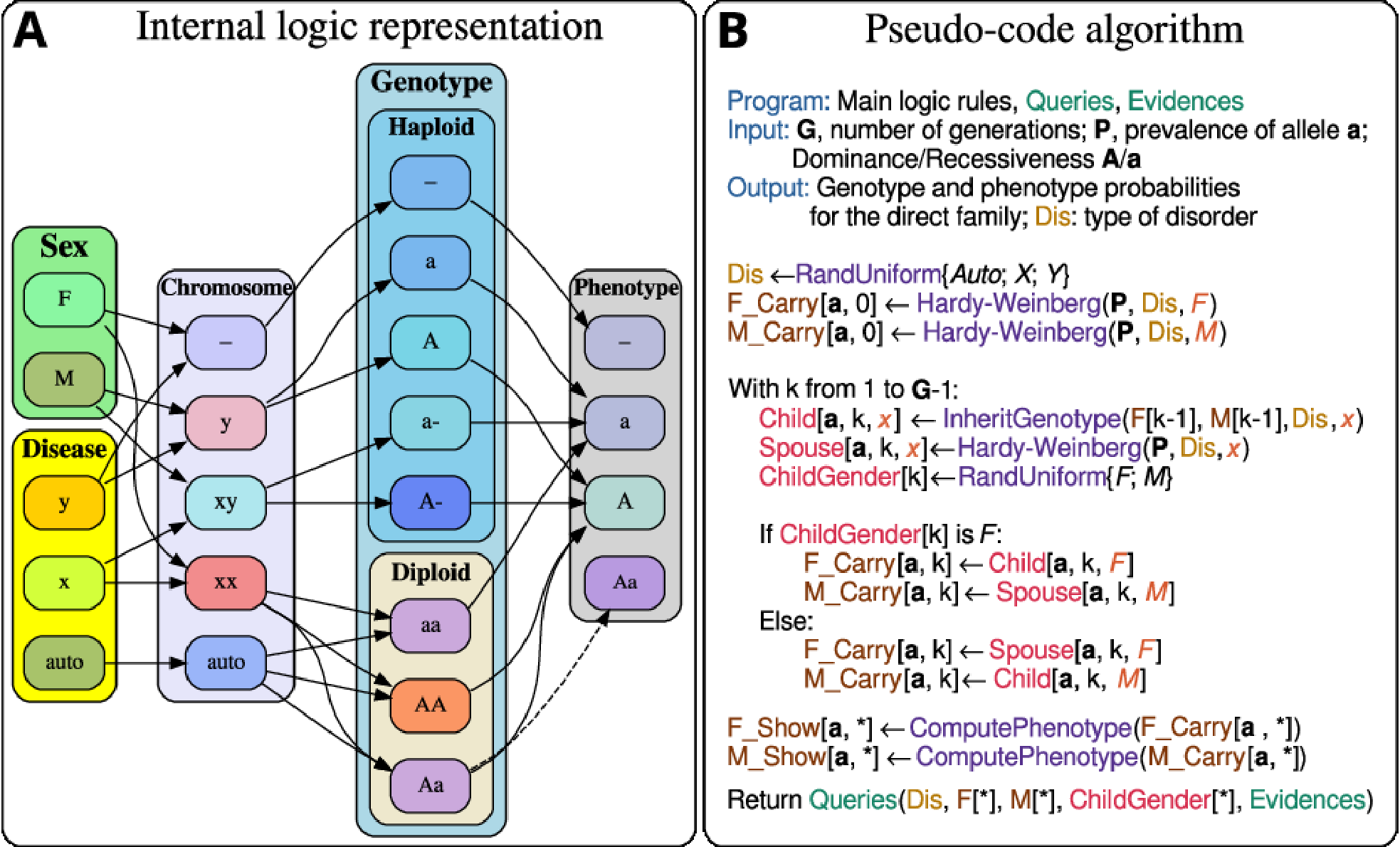}
\caption{Description of the \textit{mendelprob.pl} functioning: \textbf{A} Internal logic representation of genotypes and phenotypes, \textbf{B}: Simplified pseudo-code algorithm for genotype and phenotype probabilities transmission. Allele probabilities of direct family are inherited (\texttt{InheritGenotype}) from probabilities of parent genotypes using the internal logic representation, genotypes of spouses and last ascendants follow \texttt{Hardy-Weinberg} distribution, while disorder type and children gender -- when they are not inferred from genotypes -- assume a random uniform distribution (\texttt{RandUniform}).  Finally, \texttt{ComputePhenotype} links phenotypes and genotypes using the internal logic.}
\label{fig:methods}
\end{figure*}

The phenotype is automatically derived from the genotype using logical relationships, always assuming a penetrance (proportion of individuals with said genotype presenting said phenotype) of 100\%. Nevertheless, we provide logic rules for co-dominance (phenotype is \textit{Aa}) and incomplete penetrance (the \texttt{m\_show} and \texttt{f\_show} predicates could be weighted by a probability) if necessary. We summarize the logical relationships between phenotypes and genotypes, for each disorder type, in \hyperref[fig:methods]{Figure 1A}. 

\subsection*{Hardy-Weinberg equilibrium}

Next, using prevalence $p$ of disorder-causing alleles, we assume the well-known Hardy-Weinberg equilibrium (HWE). This will allow us to derive probabilities that any individual taken at random in a population carries a certain genotype. Hardy-Weinberg equilibrium defines $G_I$ the genotype of a random individual by the following formulas, for diploid genotypes, using $p$ the prevalence of allele $a$ in the population:

\begin{align}
    P(G_I = aa) &= p^2\\
    P(G_I = Aa) &= 2p ~(1 - p)\\
    P(G_I = AA) &= (1 - p)^2
\end{align} \vspace{1pt}

For haploid genotypes: single-allele genotypes, occurring in males in X-linked disorders and in Y-linked disorders, the formulas are simplified to having allele $a$ with a prevalence $p$ or not having allele $a$ with a prevalence $1 - p$.

\begin{align}
    ~\qquad\quad~P(G_I = a- &~|~ Dis = X,~ Sex = M) = p& \\
    ~\qquad\quad~P(G_I = A- &~|~ Dis = X,~ Sex = M) = 1 -p& \\
    ~\qquad\quad~P(G_I = a ~&~|~ Dis = Y,~ Sex = M) = p& \\
    ~\qquad\quad~P(G_I = A &~|~ Dis = Y,~ Sex = M) = 1 - p& \\
    ~\qquad\quad~P(G_I = - &~|~ Dis = Y,~ Sex = F) = 1&
\end{align} \vspace{1pt}

The Hardy-Weinberg law states that these probabilities are preserved from generation to generation, assuming that individuals in the population are randomly mating: as the number of generations increases frequencies stay constant, an equilibrium is reached \cite{thomas_statistical_2004}. In the case of our probabilistic logic program, we should assume the Hardy-Weinberg equilibrium for every generation individual whose family history is unknown. Even if the full family history were to be known, the Hardy-Weinberg law is quite a convenient tool as it could allow us to abstract a family history that is too complex. As the number of individuals and generations grow, the number of families whose genotypes should be determined would grow exponentially, so for a first approach we chose to explore only a single direct ancestry for probands, and assume Hardy-Weinberg equilibrium for the spouses and their families -- those not related to the main family. We refer to this abstraction as a \textit{direct pedigree}. 

%\begin{figure}[p]

\subsection*{Simulating disorder transmission across generations}

A probabilistic logic program composed of logic rules is used to simulate disorder transmission across generations. This constitutes the tool \textit{mendelprob.pl}, returning genotype and phenotype probabilities on \textit{direct pedigrees}, from a given  prevalence $p$ and across a number of generations $g$. Importantly, uniform random variables are chosen for modeling choice possibilities of our system. One models disorder type, between \{\texttt{auto}, \texttt{X}, \texttt{Y}\}, and, for each generation, excluding the firstmost one, random variables model the sex of the direct family descendants, between \{\texttt{M}, \texttt{F}\}. The very first generation, the first two known ancestors of the proband, inherit alleles according to HWE. Then, for each following generation, one of the individuals is a descendant of the previous generation, and the other individual is assumed distant from the family, inheriting alleles according to HWE. 

The sex of the direct family descendant, while unimportant in autosomal disorders, radically changes the transmission probabilities in X-linked and Y-linked disorders. Until this information is known, male and female genotype probabilities are thus half the probabilities of inheriting from parents, and half the probabilities from HWE. Similarly, until disorder type is known, all genotype and phenotype probabilities are \sfrac{1}{3} autosomal, \sfrac{1}{3} X-linked, \sfrac{1}{3} Y-linked. These probabilities are calculated using the queries and get reshaped by evidences given to the logic program. For instance, one evidence could be that: "the third generation descendant is a woman with pathological phenotype $a$". Such a prior knowledge excludes the possibility of a Y-linked disorder, since in Y-linked disorders phenotypes are not observed in women.

We present in \hyperref[fig:methods]{Figure 1B} a pseudo-code algorithm corresponding to the logic program, completed with queries and evidences. The logic rules for computing the inheritance of male and female alleles across generations are associated to \texttt{generation} predicates and to positive integers, \texttt{I > 0}. The probabilistic logic programming system works by defining the \texttt{generation} predicates with a rule, in the following way, where \texttt{G} in \texttt{lastgen(G)} is the number of generations minus one.

\begin{center} \tt
generation(I) :- lastgen(G), between(0, G, I), integer(I).
\end{center}

Queries and evidences are the standard way to interact with probabilistic logic programs. They are used for grounding the program, \textit{i.e.} assigning finite values to free first-order logic variables. This allows pruning of Boolean clauses irrelevant to queries and evidence out of the computation. As a result, given evidences $E$, probabilities $P(Q ~|~ E = e)$ are calculated for each query $Q$ \cite{fierens_inference_2015}. In order to inquire which genotypes and phenotypes are respectively shown by individuals of each generation, we ask the logic program the following queries:

\begin{center} \tt
\setstretch{1.2}
query(disease(\_)).\\
query(f\_carry(I, \_)) :- generation(I).\\
query(f\_show(I, \_)) :- generation(I).\\
query(m\_carry(I, \_)) :- generation(I).\\
query(m\_show(I, \_)) :- generation(I).\\
query(m\_family\_descendant(I)) :- generation(I), I > 0.\\
query(f\_family\_descendant(I)) :- generation(I), I > 0.
\end{center}

Respectively, the rules imply the following: the first query asks to estimate the probability of disease between \{\texttt{auto}, \texttt{X}, \texttt{Y}\}, the four following queries ask the probability of occurrence of the genotypes and phenotypes for the men and women at each generation $i$, the last two queries ask the probabilities that the family descendant is of a given gender for each generation $i \geq 1$. Lastly, the queries are grounded by the upper bound of \texttt{I} in \texttt{generation(I)}, defined by the predicate \texttt{lastgen(G)}. These are the queries given to \textit{mendelprob.pl} and the probabilistic logic solver for getting the probabilities shown in Results, \textit{e.g.} \autoref{tab:cpar}.

\section*{Results}

In order to illustrate the medical application of our tool, we looked at the prediction of genetic family history, also sometimes known as pedigree analysis. The first example is taken from "The practical guide to the genetic family history" by Robin Bennett \cite{bennett_practical_2011}, the second one is issued from a real American pedigree of Huntington's disease cases \cite{gusella_polymorphic_1983}. For comparing predictions of our tool, we used the state-of-the-art \textit{pedsuite} in R \cite{vigeland_pedigree_2021}.  The software suite \textit{pedsuite} allows for the elaboration and representation of pedigrees, and in particular its tool \textit{pedprobr} provides computation of probabilities for the genotypes of each family member.

\subsection*{Cystic Fibrosis}

\begin{figure}[!t]%
    \centering
    \includegraphics[width=0.6\linewidth]{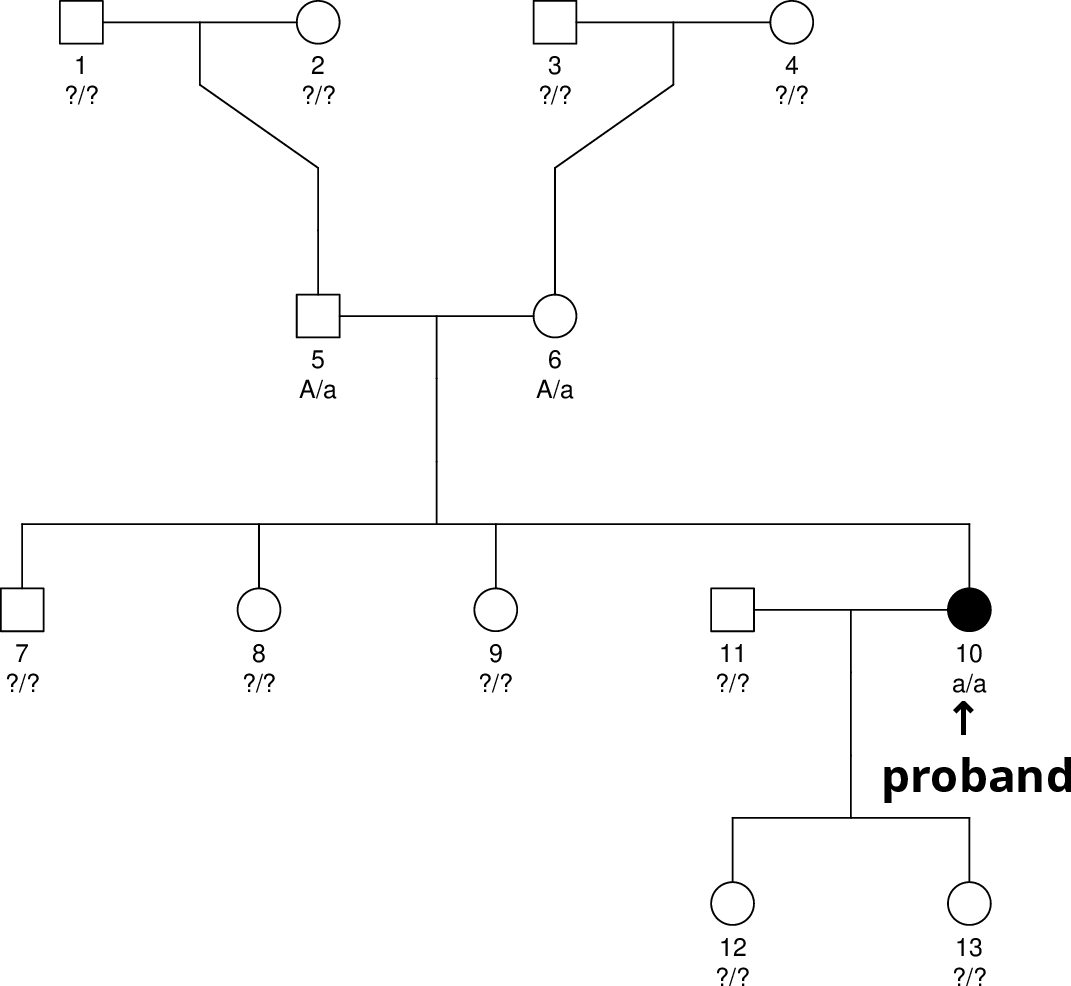}
    \caption{Pedigree of Rhonda, patient diagnosed with Cystic Fibrosis. Phenotypes marked as affected (filled black), unknown or unaffected (filled white). Pedigree constructed with \textit{pedtools} from \textit{pedsuite}.}
    \label{fig:pedigree1}
\end{figure}

Rhonda is a patient with a suspected chronic respiratory disease who recently got diagnosed with Cystic Fibrosis (CF), despite seemingly having no history of the disorder in her family. We represent her pedigree family tree in \autoref{fig:pedigree1}, as reported by Bennett \cite{bennett_practical_2011}. Rhonda got genotyped as having two mutant alleles of the CTFR gene responsible for Cystic Fibrosis. Let us model the two mutations by homozygous recessive alleles \textit{aa}. Rhonda inquires for genetic counseling. What is the probability of Rhonda transmitting the Cystic Fibrosis disease to her children?

Since we know Cystic Fibrosis's transmission is autosomal recessive, we can provide that evidence to \textit{mendelprob.pl}. By simple reasoning, we can derive that Rhonda's parents genotypes cannot be \textit{aa}, or else they would show the CF phenotype, and it cannot be \textit{AA} either, or else they would not be able to transmit the $a$ allele to Rhonda. Let us provide the following statements to \textit{mendelprob.pl}:

\begin{center} \tt
\setstretch{1.2}

prevalence("a", 1/1000). lastgen(3).\\

evidence(disease("auto")). \% Cystic Fibrosis \\

evidence(f\_carry(1, "Aa")). \% Mother of Rhonda \\ 
evidence(m\_carry(1, "Aa")). \% Father of Rhonda \\

evidence(f\_family\_descendant(2)). \% Rhonda \\ 
evidence(f\_show(2, "a")). \% Rhonda \\ 
evidence(f\_carry(2, "aa")). \% Rhonda \\

evidence(f\_family\_descendant(3)). \% Daughter of Rhonda \\
\end{center}

Here, the same probabilities are returned by our tool \textit{mendelprob.pl} and by \textit{pedprobr} using its R function \texttt{oneMarkerDistribution}. They are reported in \autoref{tab:pedigree1}. We can see that the probability of Rhonda's children (individuals 12/F3 and 13/F3) inheriting the autosomal disease (homozygous genotype \textit{aa}) from Rhonda (10/F2) and Ron (11/M2) is equal to 1/1000. This assumes that there are no cases of CF in the family of Ron, Rhonda's husband, for whom the default genotype follows HWE. We also see the tools predict possibility for Ron and Rhonda's grandparents to hold genotype $aa$.

\begin{table}[ht]
\centering
\caption{Genotype probabilities obtained with \texttt{pedprobr} and \texttt{mendelprob.pl} for the Cystic Fibrosis case of Rhonda. For column labels: numbers correspond to individuals on the pedigree tree of Figure 2, used as identifiers in \texttt{pedprobr}, while $M_i$ and $F_i$ correspond to identifiers of individuals in \texttt{mendelprob.pl}.}
\label{tab:pedigree1}
%\resizebox{0.85\textwidth}{!}{
\centering
\footnotesize
\begin{tabular}{rrrrrrrrrrr}
  \toprule
 & 1/M0 & 2/F0 & 3/M0 & 4/F0 & 5/M1 & 6/F1 & 10/F2 & 11/M2 & 12/F3 & 13/F3 \\ 
  \hline
a/a &  0.0005 &  0.0005 &  0.0005 &  0.0005 & 0 & 0  & 1 &    0.000001 & 0.001 & 0.001 \\ 
  A/a &    0.5 &    0.5 &    0.5 &    0.5 & 1 & 1 & 0 & 0.001998 & 0.999 & 0.999 \\ 
  A/A & 0.4995 & 0.4995 & 0.4995 & 0.4995 & 0 & 0 & 0 & 0.998001 &     0 &     0 \\
  \bottomrule
\end{tabular}
%}
%\vspace{8pt}
%\end{minipage}
%\vspace{-18pt}
\end{table}

Assuming Rhonda is a reliable narrator, she had never seen any other case of CF in her family. The cause of death of her grandparents from both sides are reported and none of them are related to respiratory symptoms. As well, it is fair to assume Rhonda would know if Ron was also affected by CF. Using our tool, we can verifiably exclude the case of the genotype $aa$ for Ron and for the grandparents, by specifying that their phenotype cannot be $a$. This relies on two features that \textit{pedprobr} does not currently have: expliciting phenotypes for individuals instead of genotypes, and making use of logical negation. With these features we could also avoid the previous reasoning on the parents genotypes and let the tool work automatically. Here are the new evidences:

 \begin{center} \tt
\setstretch{1.2}
evidence(not m\_show(0, "a")).  \% Either Grandfather \\
evidence(not f\_show(0, "a")). \% Either Grandmother \\
evidence(not m\_show(1, "a")). \% Father of Rhonda \\
evidence(not f\_show(1, "a")). \% Mother of Rhonda \\
evidence(not m\_show(2, "a")). \% Ron, Rhonda's husband
\end{center}

Using these evidences, genotype probabilities for the grandparents become \texttt{0.4995:carry(0,"AA")} or \texttt{0.5005:carry(0,"Aa")}. Also, for Rhonda's children, since Ron cannot carry $aa$ anymore, the probabilities for the genotype are now \texttt{0.999001:carry(3,"Aa")} and \texttt{0.000999:carry(3,"aa")}. These examples show that ProbLog can infer possible genotypes from knowledge about phenotypes alone.

Note that while the grandparents 1, 2, 3, 4 in \autoref{fig:pedigree1} are separate nodes for which we compute probabilities in \textit{pedprobr}, in \textit{mendelprob.pl} since we are only working with a single \textit{direct pedigree}, we instead model uncertainty on which grandparent the genes come from, using the probabilistic atoms \texttt{0.5:m\_family\_descendant(1)} and \texttt{0.5:f\_family\_descendant(1)}. As well, the siblings 7, 8, 9, whose genotypes are unknown and whose phenotypes bring no new information, do not belong to the direct ancestry required for a \textit{mendelprob.pl} execution. As a supplement to this article, we've included an example from Bennett's book in an appendix, available on \href{https://github.com/maxm4/mendelprob.pl/blob/main/appendix.pdf}{GitHub}, which illustrates the topic of X-linked disorders, and our tool's ability to automatically detect those.

\subsection*{Huntington's disease}

\begin{figure}
    \centering
    \includegraphics[width=0.5\linewidth]{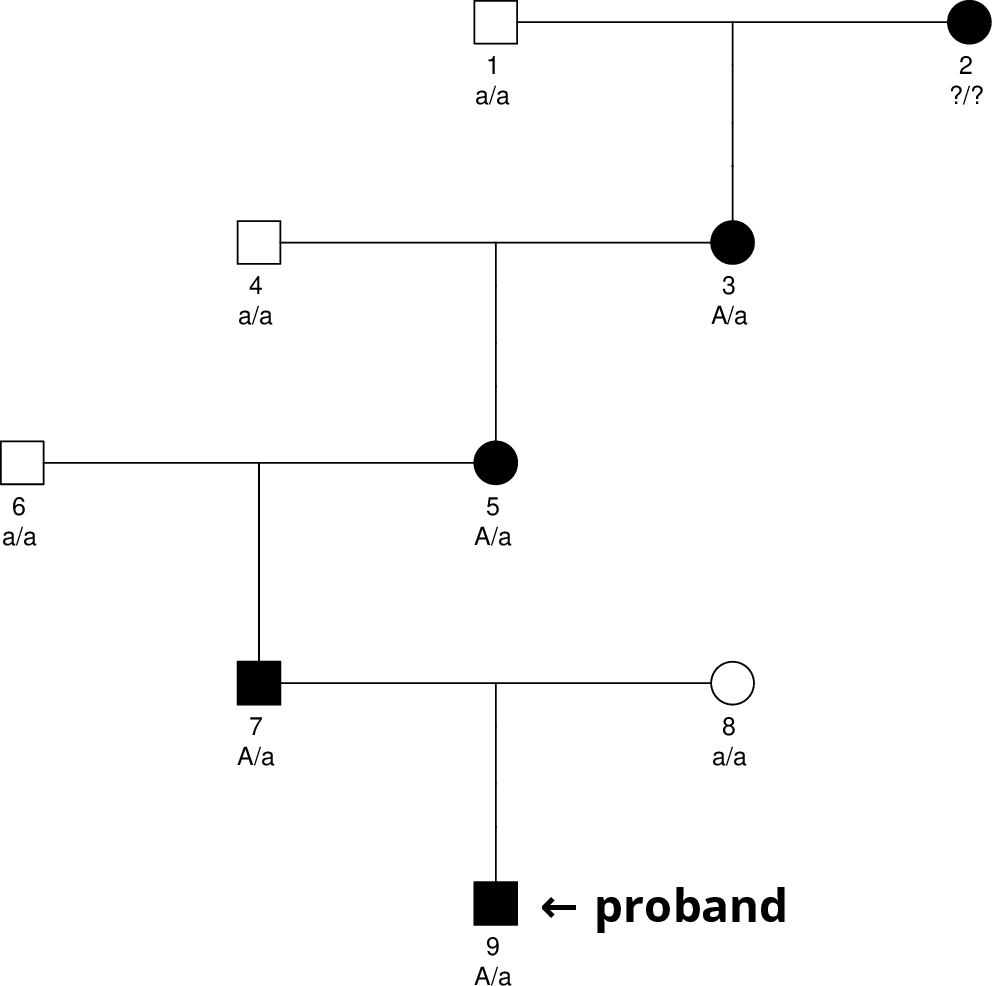}
    \caption{Pedigree of an American family with Huntington's disease, patient from the 5th generation. Phenotypes marked as affected (filled black), unknown or unaffected (filled white). Pedigree constructed with \textit{pedtools}.}
    \label{fig:pedigree3}
\end{figure}

To conclude, let us take a real example of an American family that was used to identify the gene causing Huntington's disease and some of its alleles back in 1983 \cite{gusella_polymorphic_1983}. Since our tool can again only handle a direct family and not the full family, we decided to analyze the leftmost branch of the tree, the branch leading to the sole member of the 5th generation in the article's Figure 1 \cite{gusella_polymorphic_1983}. It is represented in \autoref{fig:pedigree3}.

Huntington's disease is a well-known monogenic dominant autosomal disorder. Its prevalence is estimated at about 1/10000 according to OrphaNet \cite{pavan_clinical_2017}. Huntington's disease homozygotes (\textit{AA}) are known to be hard to identify due to dominant allele transmission \cite{wexler_homozygotes_1987}. We present the commands and evidences corresponding to phenotypes of \autoref{fig:pedigree3} below:  %\pagebreak[4]

\begin{center} \tt
\setstretch{1.2}
prevalence("a", 9999/10000). lastgen(4). \\

evidence(f\_show(0, "A")). \\
evidence(m\_show(0, "a")). \\
evidence(f\_show(1, "A")). \\
evidence(m\_show(1, "a")). \\
evidence(f\_show(2, "A")). \\ 
evidence(m\_show(2, "a")). \\
evidence(f\_show(3, "a")). \\ 
evidence(m\_show(3, "A")). \\ 
evidence(m\_show(4, "A")). \\
evidence(f\_family\_descendant(1)). \\
evidence(f\_family\_descendant(2)). \\
evidence(m\_family\_descendant(3)). \\
evidence(m\_family\_descendant(4)). \\
\end{center}

This example is pretty straightforward. From the \textit{mendelprob.pl} program, we automatically derive all of the genotypes presented in \autoref{fig:pedigree3} with probability 1, and that this is an autosomal disorder. Note that this last information was not given to ProbLog, it was inferred from impossibility of being X-linked or Y-linked. From these genotypes, we can fill in the \texttt{oneMarkerDistribution} function of \textit{pedprobr} and check whether the probabilities found match with our tool. Only a single uncertain probability remain, that of the woman at generation 0, or individual no. 2 on the pedigree. Both tools agree, its genotype probabilities are \texttt{0.9999:f\_carry(0,"Aa")} and \texttt{0.0001:f\_carry(0,"AA")}. 

\begin{table*}[ht]
\centering
\caption{Comparison of features of interest between \texttt{mendelprob.pl} and \texttt{pedprobr}, a part of \texttt{pedsuite}.} \label{tab:cpar}
{\footnotesize
\begin{tabular}{@{\extracolsep{\fill}}lp{0.1\linewidth}p{0.09\linewidth}
p{0.09\linewidth}p{0.09\linewidth}p{0.12\linewidth}p{0.12\linewidth}p{0.09\linewidth}}
\toprule
Tool          & Full \linebreak[4]pedigrees & Linkage analysis & Multi\-allelic loci & Identify \linebreak[4] disorder type & Phenotype inference \linebreak[4] from \linebreak[4] genotype & Genotype inference \linebreak[4] from \linebreak[4]phenotype & Logical \linebreak[4] reasoning  \\
\midrule
\textit{mendelprob.pl} &  \xmark  &  \xmark                 &   \xmark                 & \cmark                                          &    \cmark                               &   \cmark                                & \cmark \\                                        
\midrule
\textit{pedprobr }     & \cmark &  \cmark                & \cmark                  & \xmark                                            &  \xmark                                 &           \xmark                        &  \xmark  \\                           
\bottomrule           
\end{tabular}
}
\end{table*} %\pagebreak[4]

In conclusion, our probabilistic logic programming tool shows substantial potential for helping clinicians with constructing the genetic family history of patients. It is able to automatically derive whether a disorder is autosomal, X-linked, or Y-linked, it can be used to model recessive and dominant alleles, and it can determine genotypes across several generations with only the phenotypes in input. The differences between the \textit{pedprobr} R package and our probabilistic logic application \textit{mendelprob.pl} are reported in \autoref{tab:cpar}.

\section*{Discussion}

The \textit{mendelprob.pl} tool complements probabilistic logic descriptions of genetic problems by Blockeel and colleagues, including comparative modeling of autosomal Mendelian inheritance \cite{blockeel_probabilistic_2004} and prediction of the inheritance of multi-allelic blood type \cite{meert_cp_logic_2010}. To our knowledge, very few exhaustive open-source software for analysis of the genetic family history are freely available online. Ginsburg and collaborators listed no less than seventeen risk assessment software platforms for genetic counseling clinicians, including eight affiliated with genetic testing companies: only four are available to the public online without restrictions \cite{ginsburg_family_2019}. These are mostly commercial projects, that aren't open-source programs. For pedigree construction, it is no better, all the solutions proposed by Gordon and collaborators and by Bennett are or were commercial and through a Graphical User Interface \cite{gordon_future_2018, bennett_practical_2011}. In contrast, the drawing tool \textit{pedtools} from \textit{pedsuite} (the suite of tools from which \textit{pedprobr} is from) is open-source and in our opinion easy to use \cite{vigeland_pedigree_2021}. Vigeland, the author of \textit{pedsuite}, also developed a Graphical User Interface for his R suite: \textit{QuickPed} \cite{vigeland_quickped_2022}. Notably, \textit{pedsuite} is compatible with \textit{Familias}, another free software for probabi-\linebreak[4]listic analyses of genetic family history, used especially in forensic sciences \cite{kling_familias_2014}. Due to its original probabilistic logic programming approach, we believe our ProbLog tool \textit{mendelprob.pl} is complementary to the other tools. Therefore, we have open-sourced the tool and made it available on GitHub.

Pedigree construction \cite{gordon_future_2018} should be an important part of any genetic counsellor's \linebreak[4] analyses. Clinicians recommend three-generation pedigrees, drawn for three generations, children, \linebreak[4] parents, grandparents. This number appears as a minimum number of generations to get enough evidence from, and a fast enough procedure to be practical in medical context \cite{wattendorf_family_2005}. \linebreak[4] An important recent survey by Hussein et al. underlines that family history analysis, and especially pedigree construction, are underperformed \cite{hussein_is_2020}. Among the main complaints about pedigree construction is that it is too time-consuming \cite{hussein_is_2020, ginsburg_family_2019}. For instance Ginsburg et al. assess that the mean completion time of pedigrees with the MeTree software was 27 minutes \cite{ginsburg_family_2019}. In midst of these critics, we would like to report that our pedigree drawing experience with \textit{pedtools} from \textit{pedsuite} \cite{vigeland_pedigree_2021} was fast. Additionally, our ProbLog tool is efficient for \textit{direct pedigrees} of the recommended three generations. The advantage of using \textit{mendelprob.pl} over \textit{pedprobr} is that we can use prior knowledge about family history to infer disorder type between X-linked, Y-linked and autosomal when it is unknown, and predict genotypes when only phenotypes are known. We roughly estimate that it took us about 5-10 minutes to create a pedigree and run genotype prediction using \textit{pedtools} and \textit{mendelprob.pl}.

Gordon and colleagues hypothesized that technologies such as artificial intelligence have the \linebreak[4] potential to automate routine actions in genetic counseling, and shifting some responsibilities to the patient, including for the genetic family history \cite{gordon_future_2018}. Kearney and colleagues argued that pedigree analysis and genetic risk assessments are some of the most important tasks to automate with machine learning, but also some of the riskiest for the patient if the models were to be black-box \cite{kearney_artificial_2020}. Recently, a study of 95,166 patients involved in cancer risk assessment showed that 61,070 agreed to engage with clinical chatbots, and the authors reported a mean duration of interaction of 15 minutes with the artificial intelligence \cite{nazareth_hereditary_2021}. However, chatbots are risky to involve in such a decisive medical task as they have the ability to hallucinate. Therefore, developing formal methods in artificial intelligence for medicine is important. We hypothesize that building a natural language interface for our ProbLog tool as was envisioned for the \textit{pedtools} suite \cite{vigeland_quickped_2022}, will not only shorten patient interaction time with the tool, but also provide exact and explainable diagnoses, \textit{e.g.} \cite{vidal_explaining_2025, arias_justifications_2020}, for genetic counselors.  Further developments of our tool can be imagined beyond Mendelian inheritance, by including for example mitochondrial inheritance.

%\section{Competing interests}
%No competing interest is declared.

%\section{Author contributions statement}

%M.M. conceived and conducted the analysis. M.M. generated the figures and wrote the manuscript.

\section*{Acknowledgments}
I thank François Fages and the Lifeware team for introducing me to Prolog. Maxime Mahout reports financial support and administrative support were provided by Inria Research Centre Saclay Île-de-France. Maxime Mahout reports a relationship with Inria Research Centre Saclay Île-de-France that includes: employment. %The author declares no conflict of interests.

%\subsection*{Bibliography}

\end{document}